\documentclass[runningheads]{llncs}
\usepackage[utf8]{inputenc}

\usepackage{xcolor}
\usepackage{xspace}
\usepackage{colonequals}
\usepackage{hyperref}
\usepackage{tipa}
\usepackage{amsmath}
\usepackage{tikz}
\usetikzlibrary{arrows,automata,chains,matrix,positioning,scopes}
\usepackage{footmisc}
\usepackage{paralist}

\usepackage{listings}
\newcommand{\nodes}{\mathrm{\mathbf{N}odes}}
\newcommand{\graph}{\mathrm{\mathbf{G}}}
\newcommand{\props}{\mathit{props}}     

\newcommand{\Blank}{\ensuremath{\mathsf{Blank}\xspace}}
\newcommand{\IRI}{\ensuremath{\mathsf{IRI}\xspace}}
\newcommand{\Lit}{\ensuremath{\mathsf{Lit}\xspace}}
\newcommand{\Labels}{\ensuremath{\mathcal{L}\xspace}}

\newcommand{\sch}{\ensuremath{\mathbf{S}}\xspace}
\newcommand{\lab}{\ensuremath{\mathbf{L}}\xspace}
\newcommand{\ddef}{\ensuremath{\mathbf{def}}\xspace}
\newcommand{\tgt}{\ensuremath{\mathbf{T}}\xspace}
\newcommand{\ttgt}{\ensuremath{\mathbf{tgt}}\xspace}

\newcommand{\pat}{\ensuremath{\mathbf{P}}\xspace}
\newcommand{\plab}{\ensuremath{\mathbf{PL}}\xspace}
\newcommand{\pdef}{\ensuremath{\mathbf{pdef}}\xspace}
\newcommand{\sample}{\ensuremath{\mathbf{sample}}\xspace}

\newcommand{\subsvars}{\mathrm{sch}}

\newcommand{\svars}{\ensuremath{\mathcal{V}_\subsvars}\xspace}

\newcommand{\VOP}{[}
\newcommand{\VCL}{]}

\newcommand{\Constr}{\ensuremath{\mathsf{Constr}}\xspace}
\newcommand{\Conj}{\ensuremath{\mathsf{Conjunct}}\xspace}
\newcommand{\Choice}{\ensuremath{\mathsf{Choice}}\xspace}
\newcommand{\TConstr}{\ensuremath{\mathsf{TripleConstr}}\xspace}
\newcommand{\Slab}{\ensuremath{\mathsf{ShapeLabel}}\xspace}

\newcommand{\Pred}{\ensuremath{\mathsf{Pred}}\xspace}
\newcommand{\Prefix}{\ensuremath{\mathsf{Prefix}}\xspace}
\newcommand{\VConstr}{\ensuremath{\mathsf{ValueConstr}}\xspace} 
\newcommand{\NConstr}{\ensuremath{\mathsf{NeighbConstr}}\xspace} 
\newcommand{\VConstrs}{\ensuremath{\mathsf{ValueConstraints}}\xspace}
\newcommand{\OConstr}{\ensuremath{\mathsf{ObjectConstr}}\xspace}
\newcommand{\ListVal}{\ensuremath{\mathsf{ListValues}}\xspace}

\newcommand{\FConstr}{\ensuremath{\mathsf{UConstr}}\xspace}
\newcommand{\FNConstr}{\ensuremath{\mathsf{UNeighbConstr}}\xspace} 
\newcommand{\FTConstr}{\ensuremath{\mathsf{UTripleConstr}}\xspace}
\newcommand{\FVConstr}{\ensuremath{\mathsf{UValueConstr}}\xspace} 
\newcommand{\FCard}{\ensuremath{\mathsf{UCard}}\xspace} 
\newcommand{\FVal}{\ensuremath{\mathsf{UValue}}\xspace}

\newcommand{\Pattern}{\ensuremath{\mathsf{Pattern}}\xspace}
\newcommand{\PredDef}{\ensuremath{\mathsf{PredHolder}}\xspace}
\newcommand{\PredFilter}{\ensuremath{\mathsf{PredFilter}}\xspace}
 
\newcommand{\PNConstr}{\ensuremath{\mathsf{PNeighbConstr}}\xspace} 
\newcommand{\PTConstr}{\ensuremath{\mathsf{PTripleConstr}}\xspace}

\newcommand{\SVar}{\ensuremath{\mathsf{SVar}}\xspace}

\newcommand{\OConstrHolder}{\ensuremath{\mathsf{ObjectConstrHolder}}\xspace}

\newcommand{\MANY}{\{0;*\}}
\newcommand{\ONE}{\{1;1\}}
\newcommand{\OPT}{\{0;1\}}
\newcommand{\PLUS}{\{1;*\}}
\newcommand{\MMOP}{\{}
\newcommand{\MMCL}{\}}

\newcommand{\Fset}{F}
\newcommand{\Kset}{K}

\newcommand{\FFset}{E}

\newcommand{\iri}{\ensuremath{\mathsf{iri}}\xspace}
\newcommand{\lit}{\ensuremath{\mathsf{lit}}\xspace}
\newcommand{\nonlit}{\ensuremath{\mathsf{nonlit}}\xspace}
\newcommand{\any}{\ensuremath{\mathsf{any}}\xspace}
\newcommand{\blank}{\ensuremath{\mathsf{blank}}\xspace}
\newcommand{\xsd}{\ensuremath{\mathsf{XSD}}\xspace}

\newcommand{\clang}{SCL\xspace}
\newcommand{\uniform}{uniform\xspace}
\newcommand{\Uniform}{Uniform\xspace}

\newcommand{\fpref}[1]{#1\textrm{:}}
\newcommand{\msc}{\mathit{msc}}
\newcommand{\votecons}{\mathit{lac}}
\newcommand{\accvotes}[1]{\textit{acc}_{#1}}
\newcommand{\consensus}{\mathit{consensus}}
\newcommand{\preferred}[1]{\mathit{\pi}_{#1}}
\newcommand{\bestmatch}{\textit{best-match}}
\newcommand{\filters}{\textit{filters}}
\newcommand{\predicates}{\textit{predicates}}
\newcommand{\other}{\iri}
\newcommand{\match}{\mathit{match}}
\newcommand{\matched}{\mathit{matched}}

\newcommand{\starr}{\ensuremath{\mathsf{*}}\xspace}
\newcommand{\hole}{\_\_}

\newcommand{\shapeName}[1]{\texttt{\textless {#1}\textgreater}}
\newcommand{\lastAccess}{ retrieved on 2019/04/11}

\newcommand{\treenode}[3]{#1\\{\it\color{black!60}(#2;#3)}}
\newcommand{\treenodeinv}[3]{{\it\color{black!60}(#2;#3)}\\#1}
\newcommand{\iovkacolor}{green!10!blue!30!black}

\newcommand{\inlinecomment}[3]{#3}
\newcommand{\iovka}[1]{\inlinecomment{Iovka}{\iovkacolor}{#1}}

\newcommand{\boxcomment}[3]{}

\newcommand{\TBD}[1]{}

\begin{document}

\title{Semi Automatic Construction of ShEx and SHACL Schemas}
\titlerunning{Construction of ShEx and SHACL Schemas}

\author{Iovka Boneva\inst{1} \and
Jérémie Dusart\inst{2} \and
Daniel Fernández Álvarez \and
Jose Emilio Labra Gayo\inst{3}}

\authorrunning{I. Boneva et al.}

\institute{University of Lille, France \and
Inria, France \and
University of Oviedo, Spain}
\maketitle

\begin{abstract}

We present a method for the construction of SHACL or ShEx constraints for an existing RDF dataset.
It has two components that are used conjointly: an algorithm for automatic schema construction, and an interactive workflow for editing the schema.
The schema construction algorithm takes as input sets of sample nodes and constructs a shape constraint for every sample set.
It can be parametrized by a schema pattern that defines structural requirements for the schema to be constructed.
Schema patterns are also used to feed the algorithm with relevant information about the dataset coming from a domain expert or from some ontology.
The interactive workflow provides useful information about the dataset, shows validation results w.r.t. the schema under construction, and offers schema editing operations that combined with the schema construction algorithm allow to build a complex ShEx or SHACL schema.

  \keywords{RDF validation, SHACL, ShEx, RDF data quality, LOD usability, Wikidata}
\end{abstract}

\section{Introduction}
\label{sec:introduction}
With the growth in volume and heterogeneity of the data available in the Linked Open Data Cloud, the interest in developing unsupervised or semi-supervised methods to make this information more accessible and easily exploitable keeps growing as well.
For instance, \emph{RDF graph summarization}~\cite{cebric2018summarazing} is an important research trend that aims at extracting compact and meaningful information from an RDF graph with a specific purpose in mind such as indexing, obtaining high level descriptions of a given dataset, graph visualization, query optimization, or extracting a schema or an ontology.

The Shapes Constraint Language (SHACL)~\cite{shaclspec} is the W3C standard for defining constraints for RDF graphs.
The Shape Expressions Schemas (ShEx) language ~\cite{shexspec} has a similar purpose but is based on a slightly different set of operators.
Both proposals allow to define a set of named shapes that constrain the RDF nodes and describe their close neighbourhoods.
A set of such constraints is called a schema and provides high level information about the structure and contents of an RDF dataset.
Such schema is usually constructed with a particular application in mind.
SHACL and ShEx schemas allow to increase RDF datasets usability, as they facilitate RDF validation, understanding and querying.

More recently, several authors have used techniques close to graph summarization in order to automatically produce a SHACL~\cite{potoniec2017swift,Spahiu2018TowardsIT,melo2018automatic} or a ShEx~\cite{vandam2015rdf2graph,werkmeister2018schema,labra2018rdfshape,fernandez2018inference} schema.
Nevertheless, shapes generated using noisy or incomplete data may contain incorrect data or imprecisions.
Most of the times, the shapes should be reviewed and edited by domain experts in order to ensure they fit with the application requirements.
The usability of tools letting a proper edition of those schemas has been identified as a challenge~\cite{labra2018challenges}.
The relative youth of consolidated constraint languages for RDF data causes that this kind of tools has not been thoroughly investigated yet, but a list of desirable features has been proposed~\cite{de2018towards}.

\paragraph{Contributions}
In the current paper we describe the algorithms and the software components that are necessary for the development of an integrated, highly parametrizable and well automated tool that would allow to users with different levels of expertise to construct SHACL and ShEx schemas from data.
As a first contribution, we define an algorithm that takes as input a set of sample nodes from the RDF dataset and outputs a so called uniform shape constraint that uses only a restricted number of features of ShEx and SHACL.
The uniform constraint can either be satisfied by all the nodes in the sample, in which case it is the unique most specific such constraint.
Or it can take into account possible noise in the data and be satisfied by most sample nodes except for a small number of them consistent with a predefined error threshold.
As a next contribution, we introduce schema patterns as a general mechanism for guiding and tuning our schema construction algorithm.
A schema pattern can be provided as input to the algorithm, in which case the latter can output a complex shapes schema with nested constraints and references between shapes.
In particular, we show how to construct a schema pattern from an existing ontology so that the automatically constructed schema contains one shape for each class in the ontology and integrates the predicates' domain and range axioms.
We finally describe an interactive workflow that allows to construct an arbitrarily complex schema with the help of appropriate schema edition operators, statistical information about the data, a comprehensive visualization of validation results, and integrated calls to the algorithm for automatic schema construction.

Most of the ideas presented here are already implemented in the open source java tool \emph{ShapeDesigner} available at \href{https://gitlab.inria.fr/jdusart/shexjapp}{\url{https://gitlab.inria.fr/jdusart/shexjapp}}.

\subsubsection*{Related work}
RDF2Graph~\cite{vandam2015rdf2graph} performs structural graph summarization on knowledge bases that can be exported to RDF, OWL, XGMML or ShEx.
By default, this tool associates shapes to classes, taking into account class hierarchies.
However, some tuning is possible, allowing to define constraints for sets of nodes that do not belong to the same class.
It is limited w.r.t. expressible cardinalities but considers incoming and outgoing links in the neighbourhood of nodes.
The resulting summary can be exported to ShEx or visualized as a graphical network, but none of these options are intended to be edited or refined by the user.

RDF2Graph has been used as a base to perform shape construction in Wikidata~\cite{werkmeister2018schema}, using a publicly available tool\footnote{\url{https://tools.wmflabs.org/wd-shex-infer/} \lastAccess.}.
This tool allows to specify an initial custom set of focus nodes to be associated with a shape. 
It can relate other shapes to the initial one, being those shapes constructed w.r.t nodes of the same class.
It ignores triples in reified models, thus the schema of the metadata associated to a given piece of knowledge in Wikidata is missed.
In~\cite{fernandez2018inference} and~\cite{labra2018rdfshape} the same authors propose two tools for automatic generation of ShEx schemas, which can be partially translated to SHACL in~\cite{labra2018rdfshape}.
The tool in~\cite{fernandez2018inference} takes as input a set of classes of a Knowledge Base (KB) and produces a ShEx shape for each one computing the immediate neighbourhood of the instances of that class.
It uses statistical methods to choose the neighbourhood constraints, being able to discard the infrequent constraints and providing statistical annotations about them. 
The approach presented in~\cite{labra2018rdfshape} is independent of classes and the resulting constraints go beyond the immediate neighbourhood, but it is more limited w.r.t. cardinalities and less parametrizable.
None of those proposals define an interaction process, and they are both less parametrizable than our method w.r.t. the structure of the schemas to be constructed.
The tool presented in~\cite{potoniec2017swift} detects sets of nodes of an RDF graph that match some patterns, then combines those patterns to construct more complex OWL 2 EL class expressions and axioms.
The latter can be exported to SHACL.
The node groupings used for defining a shape are not necessarily related to the same class nor selected by the user, but they are automatically detected by the mining algorithm.

ABSTAT~\cite{principe2018abstat} is an ontology-driven tool able to produce semantic profiles of a given Knowledge Base (KB).
Those profiles gather statistics about type and cardinality of features in the immediate incoming or outgoing neighbourhood of a set of target nodes.
In~\cite{Spahiu2018TowardsIT}, ABSTAT's profiles are translated to SHACL shapes associated to the instances of some target class. 

PaTyBRED~\cite{melo2017detection} is a hybrid classifier method which relies on path and type relations for the detection of errors in KBs.
In~\cite{melo2018automatic}, the authors train decision trees using PaTyBRED to generate SHACL constraints. 
The core idea is to detect examples qualified as highly erroneous, turn them into logical conditions and use their negation as a constraint.

The novelties of our approach compared to those cited above are 
\begin{inparaenum}[(1)]
	\item we include a strategy to integrate the user in the refinement of the automatically constructed schemas; and
	\item  we introduce schema patterns as an expressive parametrization mechanism. This allows to feed the schema construction algorithm with useful information coming from an ontology, or with expert knowledge.
\end{inparaenum}

\paragraph{Organisation}
In Sect.~\ref{sec:schemas} we introduce a shapes schema language as an abstraction of ShEx and SHACL.
In Sect.~\ref{sec:simple-from-data} we define \uniform schemas as restriction of the language and explain the construction of the most specific and the largely accepted consensus \uniform constraints.
Schema patterns are defined in Sect.~\ref{sec:patterns} and the interactive workflow in Sect.~\ref{sec:interactive}.
We also discuss there the current status of the implementation of the method presented in this paper.
We finally conclude and discuss future work in Sect.~\ref{sec:conclusion}.
This submission is accompanied by a long version extended with two appendices.

\section{The Shape Constraint Language}
\label{sec:schemas}
\iovka{We assume that the reader is familiar with RDF and has at least some basic understanding of the SHACL and ShEx constraint languages.}
We start by fixing notations related to RDF graphs then present the shapes constraint language.

Denote by $\IRI$ the set of IRIs, by $\Lit$ the set of literal values, and by $\Blank$ the set of blank nodes.
Recall that an RDF graph $\graph$ is a set of triples in $(\IRI \cup \Blank) \times \IRI \times (\IRI \cup \Lit \cup \Blank)$.
If $(s,p,o)$ is a triple, then $s$ is the subject, $p$ the predicate, and $o$ the object of the triple.
A graph $\graph$ will be clear from the context all along the paper.
The elements of $\IRI \cup \Lit \cup \Blank$ that appear in subject or object position of some triple of $\graph$ are called \emph{nodes}, and the elements that appear in predicate position are called \emph{predicates} or also \emph{properties}.
  If $N$ is a set of nodes, we define $\props(N) = \{p \mid \exists (s,p,o) \in \graph \textrm{ s.t. } s \in N\}$, i.e. the set of properties of the nodes in $N$.
  The \emph{neighbourhood} of a node $n$ is the set of triples of $\graph$ with subject $n$. Node $n'$ is called a \emph{$p$-neighbour} of node $n$ if $(n,p,n')$ is a triple in $\graph$.
  \iovka{
  Throughout the paper we use IRI prefixes without defining them and assume that different prefixes (e.g. \fpref{p} and \fpref{q}) always have different definitions.}

SHACL~\cite{shaclspec} and ShEx~\cite{shexspec} are the two main languages used for defining constraints for RDF graphs. 
See~\cite{validatingRdfBook} for an introduction to these languages.
They both use the notion of \emph{shape} in order to describe or constraint the neighbourhood of nodes in RDF graphs, as illustrated in Example~\ref{ex:intuitive-shape}.

\begin{example}[The shapes \texttt{\textless Person\textgreater} and \texttt{\textless Date\textgreater}]
    \label{ex:intuitive-shape}
    The shape \texttt{\textless Person\textgreater} requires 
\begin{inparaenum}[1)]
\item an \texttt{rdf:type} link to \texttt{foaf:Person}; \item zero or more \texttt{owl:sameAs} links to some IRI;
\item a \texttt{foaf:name} link and a \texttt{foaf:familyName} link to string literals; and
\item  a \texttt{bio:birth} link to a \texttt{xsd:gYear} or a \texttt{rdgr2:dateOfBirth} link to a node that satisfies shape \texttt{\textless Date\textgreater}, but not both. 
\end{inparaenum}
On the other hand, shape \texttt{\textless Date\textgreater} requires 
\begin{inparaenum}[1)]
\item an \texttt{rdf:type} link to \texttt{time:Instant}; and
\item an \texttt{rdfs:label} that is an integer value.
\end{inparaenum}
This is how the constraint for shape \shapeName{Person} would be written in the shapes constraint language.
\begin{lstlisting}
{  rdf:type [foaf:Person] {1;1};    owl:sameAs IRI {0;*} ;                                    
   foaf:name xsd:string {1;1};      foaf:familyName xsd:string {1;1} ;
   ( bio:birth xsd:gYear {1;1} | rdgr2:dateOfBirth @<Date> {1;1} ) }
\end{lstlisting}
In Appendix~A 
    we show how these shapes are expressed in ShEx and SHACL.

In Fig.~\ref{fig:introTurtle} we provide some turtle content with a couple of nodes to be validated against the shape \texttt{\textless Person\textgreater}. \texttt{:virginia} conforms with \texttt{\textless Person\textgreater}, since it includes all the requested features regarding outgoing links.
Shapes are \emph{open} in the sense that they allow nodes to have outgoing links with predicates that are not mentioned in the shape, as for instance the property \texttt{rdgr2:placeOfBirth} of \texttt{:virginia}.
\texttt{:william} does not conform, since the information about Williams' name is provided using \texttt{schema:name} instead of \texttt{foaf:name} and \texttt{foaf:familyName}. In addition, \texttt{:william} also violates the requirement that only one among \texttt{bio:birth} or \texttt{rdgr2:dateOfBirth} links should be present.\qed
\end{example}

\begin{figure}[ht]
\begin{lstlisting}
ex:virginia a foaf:Person ;          ex:william a foaf:Person ;
  foaf:name "Virginia" ;               owl:sameAs wd:Q692 ;
  foaf:familyName "Woolf" ;            schema:name "William Shakespeare" ;
  bio:birth "1882"^^xsd:gYear ;        rdgr2:dateOfBirth bnf:1564 ;
  rdgr2:placeOfBirth "London" .        bio:birth "1564"^^xsd:gYear .
  
bnf:1564 a time:Instant ;  rdfs:label 1564 .
\end{lstlisting}
  \caption{Nodes of type \texttt{foaf:Person}}
  \label{fig:introTurtle}
\end{figure}

We now introduce the abstract Schapes Constraint Language (\clang) that containing the main features of ShEx and SHACL and has a quite straightforward translation to both of them described in Appendix~A of the long version. 
Its syntax is very similar to ShEx compact syntax.
A \emph{shape constraint} $\Constr$ is defined by the abstract syntax in Fig.~\ref{fig:syntax} using Extended Backus–Naur Form (EBNF) where the symbol `*' denotes the repetition of an element zero or more times.
We next explain the semantics of \clang.
\begin{figure}[ht]
  \centering
  \input{fig/syntax-constraints}
  \caption{Syntax of the constraint language.}
  \label{fig:syntax}
\end{figure}

A \Constr is satisfied by a node if this node and its neighbourhood satisfy both the value and the neighbourhood constraints. 
Each \VConstr restricts the value of a node to be one of the options in \VConstr, where \lit{} denotes a literal, \nonlit a non-literal, \blank a blank node, \iri any IRI, and \any imposes no value restriction, \xsd defines a required XML Schema Definition type of the value, \Prefix imposes it to be an IRI with the given prefix, and \ListVal enumerates the admissible values.
A \NConstr is satisfied by a node if its neighbourhood satisfies all the conjuncts. 
A \Choice is satisfied when just one of the \TConstr{}s composing it is satisfied.
A \TConstr is to be satisfied by the triples in the neighbourhood of the node having the specified predicate.
The number of such triples must fit within the cardinality bounds given by $min$ and $max$,  where $min$ is a natural and $max$ is either a natural or the special value '$\starr$' for \emph{unbounded} cardinality.
The objects of those triples should be nodes satisfying the specified \OConstr, which can be a simple \VConstr or another shape. 
The latter is either a reference to \Slab or a nested shape with \{\NConstr{}\}. 
Note finally that a \NConstr constrains only the triples in the neighbourhood which predicate is mentioned in one of its \TConstr. 
All other predicates can be present without restriction on their object value or cardinality.



A \emph{schema} is a pair $\sch = (\lab, \ddef)$ where $\lab \subseteq \IRI$ is a finite set of shape labels and $\ddef$ associates with every label in $\lab$ its definition which is a shape constraint.
How an RDF graph should conform to a schema is defined by a validation target.
A \emph{validation target} for schema $\sch = (\lab, \ddef)$ and graph $\graph$ is a pair $\tgt = (\lab, \ttgt)$ where $\ttgt$ associates a set of nodes of $\graph$ with every shape label from $\lab$.
We say that graph $\graph$ \emph{satisfies} schema $\sch$ with target $\tgt$ if for any shape label $L$ in $\lab$, all nodes in $\ttgt(L)$ satisfy the constraint $\ddef(L)$.
Most typically, a target is specified using a class from an ontology or a SPARQL query.
For instance, one might want to validate all nodes with class City and population greater than 1 million people against a shape \shapeName{BigCity}.
In this paper we consider non-recursive schemas only that do not allow the definition of a shape label to refer to itself neither directly nor transitively.

\section{Automatic Construction of \Uniform Constraints}
\label{sec:simple-from-data}
  Consider a fixed RDF graph $\graph$.
  In this section we address the problem of automatically constructing a shape constraint to be satisfied by a given sample set of nodes of $\graph$.
  We define \uniform constraints which enjoy the nice property that for any set of sample nodes $N$, there exists a unique most specific \uniform constraint satisfied by all the nodes in $N$ (see Sect.~\ref{sec:most-specific-constraint}).
  We also explain how to construct a \uniform constraint satisfied by most but not necessarily all sample nodes, thus allowing to take into account noise in the sample data (see Sect.~\ref{sec:majority-vote-constraint}).

  A \emph{\uniform constraint} is defined by the syntax in Fig.~\ref{fig:flat-syntax}.
  We additionally require that the \FTConstr{}s that appear in the same \FNConstr are without repeated properties, i.e. their predicates are pairwise distinct.
  Intuitively, \uniform constraints allow to enumerate the predicates to appear in the neighbourhood of a node and to specify for each predicate a constraint for its value and its cardinality among a restricted choice of four possible cardinalities.
  Remark that \uniform constraints belong to the \clang language defined in Sect.~\ref{sec:schemas}.
\begin{figure}[t]
  \centering
  \input{fig/syntax-uniform-constraints}
  \caption{Syntax of \uniform constraints.}
  \label{fig:flat-syntax}
\end{figure}

\subsection{Most Specific Constraint}
\label{sec:most-specific-constraint}
Let $\Fset$ be the set of all constraints definable by \VConstr{}s and let $\preceq$ be the (partial) ordering relation over $\Fset$ defined by $V_1 \preceq V_2$ if all nodes that satisfy $V_1$ also satisfy $V_2$.
Then for any prefix $\fpref{pr}$ and for any XSD datatype $X$ it holds
$$
\fpref{pr} \preceq \iri \preceq \nonlit \preceq \any \qquad \blank \preceq \nonlit \qquad X \preceq \lit \preceq \any
$$
Moreover, $\preceq$ is defined for prefix constraints by $\fpref{p} \preceq \fpref{q}$ if $q$ is a prefix of $p$, and also for XSD datatypes, for instance $\mathsf{int} \preceq \mathsf{decimal}$.
Finally, $\preceq$ is defined for \FVal constraints by $[r] \preceq X$ if $r$ is a literal and $X$ is its XSD type; $[r] \preceq \fpref{p}$ if $r$ is an IRI and $\fpref{p}$ is its longest prefix, and $[r] \preceq \blank$ if $r$ is a blank node.

Let $\Kset$ be the set of \uniform cardinalities, then the subset relation $\subseteq$ on intervals defines an ordering relation over $\Kset$.
It is not hard to show that
\begin{lemma}
  $(\Fset,\preceq)$ is an upper semilattice with least upper bound $\bigvee \Fset = \any$.
  $(\Kset, \subseteq)$ is a complete lattice with least upper bound $\bigvee \Kset = \MANY$ and greatest lower bound $\bigwedge \Kset = \ONE$.
\end{lemma}
Therefore, for any set of RDF nodes $R$, there exists a unique most specific \uniform value constraint satisfied by all nodes in $R$ that we denote by $\bigvee R$.
Similarly, for any set $U$ of natural numbers, there exists a unique most specific \uniform cardinality that includes $U$ and we denote it $\bigvee U$.

\begin{definition}[Most specific constraint]
  Let $N \subseteq \nodes(G)$ be a sample set of nodes. The \emph{most specific constraint} associated with $N$ is the \uniform constraint $\msc(N)$ that contains exactly one \TConstr $p~V~C$ for every predicate $p \in \props(N)$ whose value constraint $V$ and cardinality $C$ satisfy:
  \begin{itemize}
  \item $V = \bigvee N_p$ where $N_p$ is the set of $p$-neighbours of $N$;
  \item $C = \bigvee U$ where $U = \{|(n,p,n') \in \graph | \mid n \in N\}$.
  \end{itemize}
\end{definition}
It easily follows from the definition that
\begin{lemma}
  \label{lem:most-specific-flat-schema}
  For any sample $N \subseteq \nodes(\graph)$, it holds that $\graph$ satisfies the schema $(\{S\}, [S \mapsto \msc(N)])$ with target $(\{S\}, [S \mapsto N])$.
\end{lemma}
The schema defined in Lemma~\ref{lem:most-specific-flat-schema} is the most precise schema using \uniform constraints that one can construct for matching with the nodes of the sample $N$.

\subsection{Largely Accepted Consensus Constraint}
\label{sec:majority-vote-constraint}
  The largely accepted consensus constraint allows to account for potential errors in the neighbourhoods of the sample nodes.
  Imagine that most $p$-neighbours of the nodes in $N$ have an \IRI{} value, except for a very small number of them whose value is a literal.
  In this case, $\msc(N)$ will consider $\any$ as value constraint for $p$, while it is also possible that the literal $p$-neighbours are errors in the data.

  We start by presenting a decision making algorithm used in the definition of the largely accepted consensus constraint.
  Assume a non-empty set $O$ of options to choose from by a finite set of voters $W$, and let $(O, \preceq)$ be an upper semilattice with least upper bound $o_\top$, where $o \preceq o'$ indicates that if option $o$ is acceptable for given voter, then $o'$ is acceptable for him/her as well.
  Every individual $n$ votes for its preferred option $\preferred{}(n) \in O$, and let $\accvotes{\preferred{}}(o)$ be the number of voters that consider option $o$ as acceptable.
 Formally, for any total function $\preferred{}:W \to O$, we define $\accvotes{\preferred{}}(o) = \sum_{o' \preceq o} |\{n \mid \preferred{}(n) = o'\}|$.
  It is then easy to see that $\accvotes{\preferred{}}(o_\top)$ is the total number of votes, i.e. $\accvotes{\preferred{}}(o_\top) = |W|$.
  
  Let $0.5 < t \le 1$ be a threshold value.
  We define the consensus option that is ac\-cep\-table for at least a proportion $t$ of the voters.
  Let $a = \min\left\{\accvotes{\preferred{}}(o) \mid \frac{\accvotes{\preferred{}}(o)}{|W|} \ge t\right\}$
  be the least value in the range of $\accvotes{\preferred{}}$ s.t. $\frac{a}{|W|}$ is above the threshold.
  Then
  $$
  \consensus(O,\preceq,W,\preferred{},t) = \bigvee \left\{o \mid \accvotes{\preferred{}}(o) = a\right\}.
  $$
  Note that, by definition, $\consensus(O,\preceq,W,\preferred{},t)$ is uniquely defined for all semilattices $(O,\preceq)$, sets of voters $W$, vote functions $\preferred{}:W\to O$ and thresholds $t$.
  \begin{example}[Consensus]
    \label{ex:vote-illustration}
    Below on the left, consider the tree (thus also upper semilattice) $(\Fset', \preceq)$ with $\Fset' \subseteq \Fset$, where $V$ is a child of $V'$ whenever $V \preceq V'$.
    Assume a set $W$ of 20 voters and a voting function $\preferred{}:W \to \Fset'$.
    For every option $V$ we give between parentheses $(x;y)$ where $x$ is the number of direct votes (i.e. $x = |\{w \in W | \preferred{}(w) = V\}|$) and $y = \accvotes{\preferred{}}(V)$.
    Then for different values of the threshold we obtain: $\consensus(\Fset', \preceq, W, \preferred{}, \frac{17}{20}) = \iri$, $\consensus(\Fset', \preceq, W, \preferred{}, \frac{18}{20}) = \nonlit$ and $\consensus(\Fset', \preceq, W, \preferred{}, \frac{19}{20}) = \any$.
    Below on the right is presented the latitice $\Kset$, again annotated with $(x,y)$ as above for a  set $W'$ of 20 voters and a voting function $\preferred{}': W' \to \Kset$.
    Then $\consensus(\Kset, \subseteq, W', \preferred{}',\frac{17}{20}) = \bigvee\left\{\OPT,\PLUS\right\} = \MANY$.\\
    {\small
\tikzstyle{level 1}=[level distance=4em, sibling distance=4em]
\tikzstyle{level 2}=[level distance=4em, sibling distance=4em]
\tikzstyle{level 3}=[level distance=7em, sibling distance=4em]
\tikzstyle{level 4}=[level distance=8em, sibling distance=4em]
\begin{tikzpicture}[grow=right,
  every node/.style = {shape=rectangle, align=center}]
  \node {\treenode{\any}{0}{20}}
  child { node {\treenode{\nonlit}{0}{18}}
    child { node {\treenode{\blank}{1}{1}}}
    child { node {\treenode{\iri}{2}{17}}
      child { node { \treenode {\small\fpref{\texttt{http://ex.org/}}} {10} {10} } 
      }
      child { node {\treenodeinv {\small\fpref{\texttt{http://ex.com/}}}{5}{5}}}
    }
  }
  child { node {\treenodeinv{\lit}{2}{2}} }
  ;
\end{tikzpicture}
\hspace{1cm}
\begin{tikzpicture}[grow=right,node distance=5em,
  every node/.style = {shape=rectangle, align=center}]
  \node (many) {\treenode{\MANY}{0}{20}};
  \node (plus) [above right of=many] {\treenodeinv{\PLUS}{2}{18}};
  \node (opt)  [below right of=many] {\treenode{\OPT}{2}{18}};
  \node (one)  [below right of=plus] {\treenode{\ONE}{16}{16}};
  \path
  (many) edge (plus)
  (many) edge (opt)
  (plus) edge (one)
  (opt)  edge (one);
\end{tikzpicture}
}
    \qed
  \end{example}
  
  The largely accepted consensus constraint is defined using a consensus where the voters are (some of) the nodes of the sample $N$.
  For every $p \in \props(N)$, let $N_p \subseteq N$ be the set of sample nodes that have at least one $p$-neighbour and assume we are given:
  \begin{itemize}
  \item $\preferred{\Fset,p}: N_p \to \Fset$ a vote function that gives the preferred value constraint for property $p$ of every sample node that has a $p$-neighbour;
  \item $\preferred{\Kset,p}: N \to \Kset$ a vote function that gives the preferred cardinality for property $p$ of every sample node.
  \end{itemize}
\begin{definition}[Largely accepted consensus constraint]
  \label{def:maj-vote}
  Let $N \subseteq \nodes(G)$ be a sample set of nodes and $0 \le e < 0.5$ be an \emph{error rate} value.
  The \emph{largely accepted consensus constraint} associated with $N$ and assuming error rate $e$ is the \uniform constraint $\votecons_e(N)$ that contains exactly one \TConstr $p~V~C$ for every predicate $p \in \props(N)$ and it has value constraint $V = \consensus(\Fset, \preceq, N_p, \preferred{\Fset,p}, 1-e)$, and cardinality $C = \consensus(\Kset, \subseteq, N, \preferred{\Kset,p}, 1-e)$. \qed
\end{definition}
The error rate value $e$ is intuitively a bound on the proportion of sample nodes that could have expressed an incorrect\footnote{Incorrect w.r.t. a supposedly existing correct \uniform constraint that the user aims at.} vote due to noise in the data.
It is not hard to see that $\votecons_0(N) = \msc(N)$.

Finally we define the vote functions $\preferred{\Fset,p}$ and $\preferred{\Kset,p}$.
For any node $n \in N$ and any $p \in \props(N)$, $\preferred{\Kset,p}(n)$ is defined as follows, where $m$ is the number of $p$-neighbours of $n$:
$$
\preferred{\Kset,p}(n) = \begin{cases}
  \OPT & \textrm{ if } m = 0\\
  \ONE & \textrm{ if } m = 1\\
  \PLUS & \textrm{ if } m > 1
\end{cases} 
$$
As for $\preferred{\Fset,p}$, it is itself defined as a consensus: $\preferred{\Fset,p}(n) = \consensus(\Fset, \preceq, M, \bestmatch, 1-e')$ where $0 \le e' < 0.5$ is an error rate that might be different from $e$, the voters $M = \{n' \mid (n,p,n') \in \graph\}$ are all $p$-neighbours of $n$, and the voting function $\bestmatch:M \to \Fset$ is defined by $\bestmatch(n') = [n']$ (i.e. the singleton \ListVal  containing $n'$).

\paragraph{Discussion}
The $\msc$ and $\votecons$ constraints are implicitly parametrized by $\Fset$ that could be replaced by any upper semilattice.
For instance, $\Fset = \{\any, \blank, \lit\}$ might be sufficient for some use cases.
We choose to present this particular value for $\Fset$ because it is useful in practice.

\section{Schema patterns}
\label{sec:patterns}
\begin{figure}[h]
\input{fig/fig-auckland}
\caption{Extract of the Wikidata entry for Auckland}
\label{fig:auckland}
\end{figure}
The main limitation of \uniform constraints is that they do not allow for nested neighbourhood constraints or references to other shapes.
A possible alternative could be to construct a schema up to some fixed nesting depth given as parameter \cite{werkmeister2018schema,labra2018rdfshape} but the same nesting depth is seldom appropriate to all shapes.
We propose to use schema patterns instead.
They allow for a fine tuning of automatic schema construction by describing a general form for the target schema, but can also be used simply to restrict the namespaces of predicates of interest, or to give a particular role to some values such as the value of \texttt{rdf:type}.

We start by a motivating example based on Wikidata that we use for an informal description of schema patterns in Sect.~\ref{sec:patterns-informal}.
Then in Sec~\ref{sec:patterns-formal} we give the syntax of schema patterns.

\subsection{Informal Definition}
\label{sec:patterns-informal}
  Fig.~\ref{fig:auckland} presents an extract of the Wikidata entry for the city of Auckland\footnote{From {\small\url{https://www.wikidata.org/entity/Q37100.ttl}} retrieved on 2019/03/27.}\footnote{\label{fn-props}restricted to the properties \texttt{rdf:type}, \texttt{P17} and \texttt{P6}.}.
  Entities in Wikidata are encoded by identifiers that start with \texttt{Q}, and properties by identifiers that start with \texttt{P}.
  The entry in Fig.~\ref{fig:auckland} includes a statements for the property country (\texttt{P17}) with value New Zealand (\texttt{Q664}), and two statements for the property head of government (\texttt{P6}) with values Phil Goff (\texttt{Q597909}) and Len Brown (\texttt{Q4116955}).
  The most significant value of each property is given as a direct statements (ds) (prefix \texttt{wdt:}), while richer information is presented in a qualified \texttt{wikibase:Statement} in which the claim is accompanied by context or scope information (under which conditions the claim holds), as well es meta information such as provenance.
  Note that the structures of such statements are similar.
  They usually contain property statements (\texttt{ps:}), qualifiers (\texttt{pq:}) and provenance information.
 
In Fig.~\ref{fig:msc-cities} is presented the $\msc(N)$ constraint where the sample $N$ contains the Wikidata entries for Auckland, Monterey, Vienna and Kobe\footref{fn-props}, which is clearly too poor to represent the actual structure of the data.
\begin{figure}[t]
  \input{fig/fig-msc-cities}
  \caption{Most specific constraint for a sample of Wikidata cities}
  \label{fig:msc-cities}
\end{figure}
A more appropriate schema $\sch_\textrm{city}$ for cities in Wikidata\footref{fn-props} is given in Fig.~\ref{fig:wikidata-desired-schema}.
It accounts for the possible qualifiers of the different properties.
\begin{figure}[t]
  \input{fig/fig-wikidata-desired-schema}
 \caption{Schema $\sch_\textrm{city}$ for Wikidata cities}
 \label{fig:wikidata-desired-schema}
\end{figure}

A schema similar to $\sch_{\textrm{city}}$ can be automatically constructed using the schema pattern $\pat_{\textrm{city}}$ in Fig.~\ref{fig:pattern-cities} as input of the constraint construction algorithm.
A schema pattern is similar to a schema but it omits cardinalities, it allows predicate filters (\texttt{p:}, \iri, \texttt{ps:}, \ldots) instead of predicates, it replaces value constraints by placeholders $\hole$ or $[\hole]$, and it replaces some shape names by variables ($Y$).

The pattern $\pat_{\textrm{city}}$ defines one non-variable shape label \texttt{<City>}, therefore it will be used with one set of sample nodes, say $N_{\textrm{city}}$, in order to construct a shapes schema.
Denote $P_{\textrm{city}}$ the pattern for \texttt{<City>} and assume that $N_{\textrm{city}}$ contains the single node \texttt{wd:Q37100} from Fig.~\ref{fig:auckland}.
We start by matching all properties $q$ in $\props(N_{\textrm{city}})$ with the predicate filters that appear in the $P_{\textrm{city}}$, and every such property $q$ is matched with the most specific possible filter.
That is \texttt{p:P17} and \texttt{p:P6} are matched with \texttt{p:} and we write $\matched(\texttt{p:}) = \{\texttt{p:P16}, \texttt{p:6}\}$, while $\matched(\texttt{iri}) =\{\texttt{wdt:P17}, \texttt{wdt:P6}, \texttt{rdf:type}\}$.
Now every property in $\matched(\texttt{p:})$ yields a triple constraint, and the pattern requires that its object is \texttt{@Y}, a shape reference to a freshly created shape label.
We can see that schema $\sch_{\textrm{city}}$ in Fig.~\ref{fig:wikidata-desired-schema} contains indeed two triple constraints \texttt{p:P17 @<Y\_P17>} and \texttt{p:P6 @<Y\_P6>}.
The cardinalities of these triple constraints are not relevant here as they were obtained by considering a bigger sample set $N_{\textrm{city}}$.
In the general case, any property $q \in \matched(\texttt{p:})$ defines its corresponding sample $N_q = \{n' \mid \exists n \in N_{\textrm{city}}. (n,q,n') \in \graph\}$ composed of the $q$-neighbours of the nodes in $N_{\textrm{city}}$.
This sample is used to determine the cardinality of the triple constraint on the one hand, and as a sample for the definition of shape label \texttt{<Y\_$q$>} on the other hand.
Similarly, every property in $\matched(\iri)$ yields a triple constraint which object constraint is a value constraint, as indicated by the placeholder \hole.
Now regarding the pattern $P_Y$ for $Y$, we already saw that it defines two different shapes as defined by $\matched(\texttt{p:})$, and each of these has its associated sample.
Remark that $P_Y$ does not contain filter \iri, therefore some of the properties of the sample nodes might match none of the filters.
These properties will not be considered.
For instance, property \texttt{prop:wasDerivedFrom} appears in the sample nodes in Fig.~\ref{fig:auckland}, but not in the schema $\sch_{\textrm{city}}$.
We finally explain the triple constraint \texttt{rdf:type [\hole]} in $P_Y$.
Its predicate holder is not a filter but a property, so it will be matched only by this same property.
Its object constraint is the placeholder [\hole] indicating that we want to construct a \ListVal constraint for the set of values.
In the example this is the singleton list that contains \texttt{wikibase:Statement}.
\begin{figure}[th]
  {\small\input{fig/fig-pattern-cities}}  
  \caption{Schema pattern $\pat_{\mathrm{city}}$ for Wikidata cities}
  \label{fig:pattern-cities}
\end{figure}

As we saw, schema patterns are particularly fitted for the Wikidata dataset as it strongly relies on reification.
But they are also useful for all datasets that use reification \cite{frey2017evaluation} or repetitive local structure.
Simpler schema patterns can also be used for e.g. restraining the properties of interest.

\subsection{Formal Definition}
\label{sec:patterns-formal}
Assume a countable set $\svars$ of shape label variables ranging over $X,Y,\ldots$.
A \emph{constraint pattern} (or pattern for short) is defined by the syntax in Fig.~\ref{fig:syntax-patterns}.
\begin{figure}[t]
  \centering
  \input{fig/syntax-patterns}
  \caption{Syntax of constraint patterns.}
  \label{fig:syntax-patterns}
\end{figure}
A \emph{schema pattern} is a triple $\pat = (\plab, \pdef, \sample)$ where:
\begin{itemize}
\item $\plab \subseteq \Labels \cup \svars$ is a finite set of shape labels and shape label variables;
\item with every label or variable in $\plab$, $\pdef$ associates a shape pattern that uses shape variables from $\plab\cap\svars$ only;
\item $\sample$ associates a set of nodes from $\graph$ with every label in $\plab \cap \Labels$.
\end{itemize}
A schema pattern defines a most specific \uniform schema (or a largely accepted consensus schema) which definition is given in Appendix~B
of the long version due to space limitations.

\subsection{Patterns for Ontologies}
\label{sec:ontology}
We show here how a schema pattern can be used to encode some of the information available in an existing ontology that might be relevant for the schema construction.
Consider the toy RDFSchema ontology $R$ that defines  three classes \texttt{:Human}, \texttt{:Teacher} and \texttt{:Subject} with \texttt{:Teacher} subclass of \texttt{:Human}, and three properties \texttt{:name}, \texttt{:teaches}, \texttt{:description} with axioms
\begin{lstlisting}
  :name    rdfs:domain :Human.    :description rdfs:domain :Subject.
  :teaches rdfs:domain :Teacher.  :teaches     rdfs:range  :Subject.
\end{lstlisting}
We can automatically retrieve the following schema pattern 
\begin{lstlisting}
  <Human>   { a [ %\hole% ] ;  :name %\hole% }
  <Teacher> { a [ %\hole% ] ;  :name %\hole% ;  :teaches %\hole% }
  <Subject> { a [ %\hole% ] ;  :description %\hole% }    
\end{lstlisting}
with $\ttgt(\texttt{<Human>})$ defined by the SPARQL query \texttt{SELECT ?x WHERE ?x a :Human.} and similarly for \texttt{<Teacher>} and \texttt{<Subject>} shapes.
The shapes schema constructed with this pattern will contain only the properties of the ontology vocabulary attached to their respective domains.

\section{Interactive Improvement of the Schema}
\label{sec:interactive}
We are convinced that an optimal shapes schema, i.e. a schema that encodes correctly the requirements for a particular application, cannot be obtained by a fully automatic process.
Automatically constructed schemas are necessarily simple: if we allowed the choice operator or repeated properties\footnote{that is, allow two triple constraints with same predicate in the same neighbourhood constraint} in \uniform constraints, then a unique most specific constraint would not exist.
Schema patterns (or other forms of parametrization) can bring a partial solution, but they may require user expertise and good knowledge about the data, which is often the reason why we want to build a schema at the first place: a chicken-and-egg situation.

Our solution is to let the user build the schema but within the interactive workflow described hereafter that gives access to automatic schema construction and to useful information about the data, and provides appropriate feedback.

Assume an RDF graph $\graph$ for which we want to build a schema.
The interactive workflow maintains a schema $\sch = (\lab, \ddef)$ together with associated validation target $\tgt = (\lab, \ttgt)$ and schema pattern $\pat = (\plab, \pdef, \sample)$ with $\sample = \ttgt$.A target $\ttgt(L)$ is given as a tuple $(Q, N^+, N^-)$ of a SPARQL query $Q$ to be executed against $\graph$ and sets of nodes $N^+$ and $N^-$, to be added, resp. removed from the result of query $Q$.
The schema and the targets can be modified using the following editing operations:
\begin{itemize}
\item modify $\ddef(L)$ for shape label $L$ using these actions:
  \begin{itemize}
  \item add or remove a \Conj or \VConstrs;
  \item group two or more triple constraints into a \Choice, or split a choice;
  \item split a \TConstr of the form $p ~ V ~ C$ in two \TConstr{}s $p ~ V_1 ~ C_1$ and $p ~ V_2 ~ C_2$ having the same predicate, or regroup two such triple constraints;
  \item change the cardinality or the \OConstr of some triple constraints with predicate $p$.
    In particular, the object constraint can be replaced by the automatically constructed $\{ \msc(N') \}$ or $\{ \votecons(N') \}$, where $N'$ is the set of $p$-neighbours of $\ttgt(L)$;
  \end{itemize}
\item modify $\ttgt(L)$ for shape label $L$.
  This action is triggered automatically when some triple constraint is changed and becomes of the form $p ~ @L ~ C$ (with shape reference to $L$).
  Then the set $N'$ of all IRI $p$-neighbours of $\ttgt(L)$ is added to $\ttgt(L')$.
    In the latter case, the \NConstr $M$ is automatically constructed as $\msc(N')$ or $\votecons(N')$.
\item add a new shape label $L$ to $\lab$ with corresponding target $\ttgt(L) = N$ and pattern $\pdef(L)$, which triggers the automatic construction of $\msc(N)$ or $\votecons(N)$ using $\pdef(L)$ as pattern and assigns the result to $\ddef(L)$.
\end{itemize}
These operations allow to construct an arbitrarily complex \clang schema starting e.g. from an automatically constructed one.

At any time the user can visualize the result of validating $\graph$ against $\sch$ with target $\tgt$.
Additionally, the user has access to the following information:
\begin{itemize}
\item for any shape label $L$ and triple constraint $p ~ V ~ C$ that occurs in $\ddef(L)$ or in $\pdef(L)$, and denoting $N = \ttgt(L)$:
  \begin{itemize}
  \item the number of nodes in $N$ that have $p$ as predicate;
  \item the minimal, maximal and average number of occurrences of predicate $p$ in the neighbourhood of $N$;
  \item the lattice of \uniform value constraints $(\Fset,\preceq)$ annotated with $\accvotes{\preferred{F}}(V')$ for every $V' \in \Fset$ (see Sect.~\ref{sec:most-specific-constraint});
  \end{itemize}
\item for any pair of predicates $p,p'$, the number of nodes from $N$ in which $p$ and $p'$ co-occur; this is useful to detect possible \Choice{}s.
\end{itemize}

\subsubsection{Implementation}
We have already implemented several components of our method, some of which are illustrated in the video attached to the submission:
\begin{itemize}
\item the algorithm for automatic construction of the most specific schema with only limited support of schema patterns tailored for Wikidata (video),
\item simple visualisation of statistics about the data,
\item enhanced visualisation of validation results that allows to navigate synchronously in the schema and in the graph to understand why a given node is valid or not w.r.t. a given constraint (video).
\end{itemize}
Currently our implementation is based on ShEx validation.
In order to complete the implementation we need to:
\begin{inparaenum}[(1)]
\item complete the implementation of schema patterns;
\item add support for SHACL validation;
\item provide a nice visualisation of the information about the data; and
\item integrate the different components into a single application.
\end{inparaenum}

\begin{figure}[ht]
  \includegraphics[width=\linewidth]{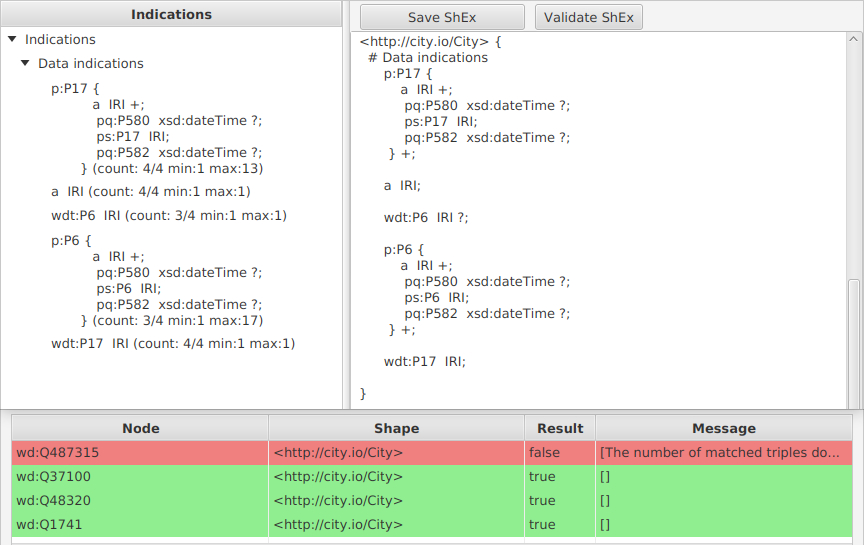}
  \caption{Screenshot of some parts of the tool}
  \label{fig:app}
\end{figure}

Fig.~\ref{fig:app} shows a consolidated view of three of the already implemented components.
We consider a sample with four Wikidata entries: Auckland, Monterey, Vienna and Kobe (restricted to a part of their predicates only).
The top right panel contains the schema that was automatically constructed using a schema pattern similar to the one presented in Fig.~\ref{fig:pattern-cities}.
We changed a single cardinality in that schema that produced a validation error.
In the bottom panel we see the list of validated nodes on which the error is highlighted.
The top left panel shows part of the information provided to the user, i.e. the list of the properties of sample nodes and their frequencies of occurrence.




\section{Conclusion and Future Work}
\label{sec:conclusion}
We have presented a method that allows to define ShEx or SHACL schemas for existing datasets.
The construction of simple (\uniform) schemas is fully automatic and is accessible to novice users, while experts can supply their knowledge about the data using schema patterns as an expressive parametrization mechanism, and further edit the resulting schema.
A prototype implementation of the method allows to extract and refine ShEx schemas for Wikidata.
We are currently extending the implementation in order to integrate the still missing features: full support for schema patterns, translation to SHACL and integrated SHACL validation, a more readable representation of the statistics about the data using GUI widgets.
As a future work we are planning to enhance schema patterns in order to capture richer information that comes from ontologies such as cardinalities, and also to allow to describe more complex references between shapes.

\bibliographystyle{splncs04}


\appendix
\section{The Shape Constraint Language, SHACL and ShEx}
\label{sec:app-examples}
We explain here how \clang is translated to SHACL and to ShEx, then we explain the small differences in semantics depending on the target language.

\paragraph{Translation to SHACL}
\clang's \VConstr{}s are represented using \texttt{sh:nodeKind} (for \lit, \nonlit, \blank, \iri), using \texttt{sh:datatype} for \xsd, using \texttt{sh:pattern} for \Prefix, and using \texttt{sh:in} for \ListVal.
A \NConstr yields a \texttt{sh:shape} that is a conjunction (\texttt{sh:and}) of the encodings of its \Conj{}s.
The \Choice operator can be represented by \texttt{sh:or} or by \texttt{sh:xone} (which one to use is a parameter of the translation).
\TConstr is encoded using \texttt{sh:property} with \texttt{sh:path} constraint for its predicate, while its \OConstr is represented by \texttt{sh:node} if it is a shape reference or a nested \NConstr, and by the encoding of the \VConstr otherwise.
Cardinality is represented by \texttt{sh:minInclusive} and \texttt{sh:maxInclusive} as expected.
Additionally, if two or more triple constraints in the same neighbourhood constraint have the same predicate, then all these triple constraints are encoded using \texttt{sh:qualifiedValueShape}, and their cardinalities using \texttt{sh:qualifiedMinCount} and \texttt{sh:qualifiedMaxCount}.
Finally, SHACL shapes have associated target declarations that are not provided by the \clang schema but by its accompanying validation target.
In Fig.~\ref{fig:introShacl} we give as example the shapes \texttt{\textless Person\textgreater} and  \texttt{\textless Date\textgreater} from Sect.~\ref{sec:schemas} written in SHACL.
\begin{figure}[ht]
\begin{lstlisting}
 <Person>  a sh:NodeShape ;
     sh:property [ sh:path owl:sameAs ;   sh:nodeKind sh:IRI ;
          sh:minCount 0  ] ;
     sh:property [ sh:path foaf:name ;   sh:datatype xsd:string ;
          sh:minCount 1 ;   sh:minCount 1 ] ;
     sh:property [ sh:path foaf:familyName ;   sh:datatype xsd:string ;
          sh:minCount 1 ;   sh:minCount 1 ] ;
     sh:property [ sh:path rdf:type;   sh:hasValue  foaf:Person ;
          sh:minCount 1; sh:maxCount 1 ]  ;
     sh:xone ( [ sh:property [ sh:path bio:birth ;   sh:datatype xsd:gYear ;
                    sh:minCount 1 ;  sh:maxCount 1 ] ]
               [ sh:property [ sh:path rdgr2:dateOfBirth ;   sh:nodeKind sh:IRI ;
                    sh:minCount 1 ;   sh:maxCount 1 ; ] ] ) ;
     sh:closed false ;   sh:targetClass foaf:Person . 
<Date>  a sh:NodeShape ;
     sh:property [ sh:minCount 1 ;   sh:maxCount 1 ;
          sh:path rdf:type ;   sh:hasValue time:Instant ] ;
     sh:property [ sh:minCount 1 ;   sh:maxCount 1 ;
                   sh:path rdfs:label ;   sh:datatype xsd:int ] ;
     sh:closed false;   sh:targetClass time:Instant .
\end{lstlisting}
  \caption{Shape \texttt{\textless Person\textgreater} in SHACL}
  \label{fig:introShacl}
\end{figure}

\paragraph{Translation to ShEx}
The syntax of \clang is based on the compact syntax of ShEx, therefore the translation to ShEx is almost straightforward, except for \Conj that becomes \textit{EachOf} also denoted $;$ in the compact syntax, and \Choice that becomes \textit{OneOf} denoted $\mid$.
Additionally, cardinality \ONE{} is omitted, cardinaity \MANY{} becomes \texttt{*}, \PLUS{} becomes \texttt{+} and \OPT{} becomes \texttt{?}, while any other cardinality remains as it is.
In Fig.~\ref{fig:introShex} we represent the translation to ShEx of shapes \texttt{\textless Person\textgreater} and  \texttt{\textless Date\textgreater} from Sect.~\ref{sec:schemas}.
\begin{figure}[ht]
\begin{lstlisting}
<Person> { rdf:type [foaf:person] ;
           owl:sameAs IRI * ;
           foaf:name xsd:string ;
           foaf:familyName xsd:string ;
           ( bio:birth xsd:gYear | rdgr2:dateOfBirth @<Date> ) }
<Date>   { rdf:type [time:Instant] ; rdfs:label xsd:int }
\end{lstlisting}
  \caption{Shape \texttt{\textless Person\textgreater} in ShEx}
    \label{fig:introShex}
\end{figure}

\paragraph{Discussion} The semantics of a \clang schema can be different if it is translated to ShEx or to SHACL.
The two differences come from the choice operator and from the use of different triple constraints with the same property in the same neighbourhood constraint.
This is not an issue for our approach because we make the hypothesis that the user constructs a schema for one of the two languages, not for both, depending on the application she or he has in mind and on her/his production suite.
The validation results in the interactive tool are computed using the translation to the chosen target schema language, whichever it is.
When the desired abstract \clang schema is ready, it is exported and can be used in production.

\section{Formal Definition of Shape Patterns}
\label{sec:app-patterns}
A schema pattern is defined by the syntax in Fig.~\ref{fig:syntax-patterns}.
We additionally require that (1) no \PredFilter is used more than once in the same constraint pattern, and (2) every shape label variable $Y \in \svars$ is used exactly once in the range of $\pdef$.

Denote by $\FFset$ the set of \PredFilter{}s of constraint patterns.
Remark that $\FFset$ is a subset of the set $\Fset$ of value constraints defined in Sect.~\ref{sec:simple-from-data}, and that $(\FFset, \preceq)$ is an upper semilattice with least upper bound $\bigvee \FFset = \other$.
For any \PNConstr $P$, denote $\FFset(P)$ the set of predicate filters that appear in the direct \PTConstr{}s of $P$ (that is, excluding those that appear in some nested \PNConstr).
Because $\FFset(P) \subseteq \FFset$ and $(\FFset,\preceq)$ is a lattice, it holds that for every property $q$ either no element of $\FFset(P)$ matches $q$ and we write $\match(P,q) = \bot$, or there is unique most precise $f$ in $\FFset(P)$ s.t. $[q] \preceq f$ and we write $\match(P,q) = f$.

For a property $q$, denote $\FFset(P)_{q\preceq}$ the set of filters in $\FFset(P)$ that match $q$, i.e. $\FFset(P)_{q\preceq} = \{f \in \FFset(P) \mid [q] \preceq f\}$.
For instance, if $P$ is $\pdef(Y)$ from Fig.~\ref{fig:pattern-cities} then $\fpref{ps} \in \FFset(P)$ and $\fpref{ps} \in \FFset(P)_{\texttt{ps:foo}\preceq}$. 

Fix a pattern schema $\pat = (\plab, \pdef, \sample)$ for the sequel.
Its associated most specific schema with \uniform constraints $\msc(\pat)$ is defined below.
We point out first that $\msc(\pat)$ is not a \uniform schema as it allows nested neighbourhood constraints and shape references, but likewise \uniform schemas, its neighbourhood constraints satisfy do not use choice or repeated properties.
So let $\msc(pat) = (\lab, \ddef)$, we explain now which are the shape labels in $\lab$ and which are their definitions.

For every shape label $L \in \plab$, $L$ is in $\lab$ and its its definition $P = \pdef(L)$ together with its sample $N = \sample(L)$ define a \NConstr $S = \msc(P,N)$.
$S$ is a conjunct of \TConstr{}s (without \Choice{}s).
Denote by $\filters(P)$, resp. by $\predicates(P)$, the set of \Pred{}icates, resp. of \PredFilter{s}, that occur in the \PTConstr{}s of $P$.
For instance, $\filters(\pdef(Y)) = \{\texttt{ps:},\, \texttt{pq:}\}$ and $\predicates(\pdef(Y)) = \{\texttt{rdf:type}\}$ for the pattern for shape $Y$ in Fig.~\ref{fig:pattern-cities}.
Then let $\props(P,N)$ be the set that contains exactly the properties that apper in $P$ and all properties that appear in the neighbourhood of $N$ and match some of the filters in $P$.
$$
\props(P,N) = \predicates(P) \cup \left\{ q \in \props(N) \mid \match(P,q) \neq \bot \right\}.
$$
Finally, for every predicate $q \in \props(P,N)$, its associated sample $s(q)$ is the set of $q$-neighbors of $N$.

Each $q \in \props(P,N)$ yields a triple constraint $q ~ O ~ C$ which object constraint $O$ and cardinality constraint $C$ are defined as follows.
Let $H_q$ be the object constranint holder for $q$ by $P$.
That is, if $q \in \predicates(P)$, then $H_q$ is the object constraint holder associated with $q$ in $P$.
Otherwize, $H_q$ is the object constraint holder associated with $\match(P,q)$, the most precise predicate filter that matches $q$. 

Now, the object constraint $O$ is, depending on the form of $H_q$, and denoting $N_q$ the set of $q$-neighbours of $N$:
\begin{itemize}
\item If $H_q = \hole$ (a value constraint holder), then $O = \bigvee N_q$.
\item If $H_q = [\hole]$ (a list of values holder), then $O$ is the list that contains exactly the values in $N_q$.
\item If $H_q = \{M\}$ (a nested neighbourhood constraint pattern), then $M = \msc(P,N_q)$.
\item If $H_q$ is of the form $@X$ (shape reference), then we create a fresh shape label \texttt{<X\_q>} and $O = @\texttt{<X\_q>}$.
We add \texttt{<X\_q>} to $\lab$ and its definition $\ddef(\texttt{<X\_q>}) = \msc(\pdef(X), N_q)$.
\end{itemize}

Because the pattern $\pat$ is not recursive (requirement (2) above), this procedure terminates and construcs a shapes schema.

We finally point out that schema patterns might suffer of the problem of ``shrinking samples''.
Even if the size of $\sample(L)$ for some $L \in \plab$ is sufficient, its nested neighbourhood constraints are constructed using samples of $q$-neighbours of the nodes in $N$ for the appropriate predicate $q$.
If predicate $q$ occurs only a few times in the neighbourhood of $N$, then the corresponding sample can be very small.
In our implementation we warn the user if such situation occurs, and allow her to add additional nodes to that sample without necessarily augmenting $\sample(L)$ (for efficiency and readability reasons).

\end{document}